\documentclass[twocolumn]{revtex4}

\usepackage{graphicx}                                                                            
\graphicspath{{figures/}}

\usepackage{color}                                                                                 
\usepackage{float}                                                                                  
\usepackage{wrapfig}                                                                             

\usepackage[parfill]{parskip}                                                                  

\usepackage{amstext,amsmath,amssymb,amsfonts,braket}                 

\usepackage{hyperref}

\newcommand{\beq}{\begin{equation}}
\newcommand{\eeq}{\end{equation}}

\newcommand{\beqq}{\begin{equation*}}
\newcommand{\eeqq}{\end{equation*}}

\newcommand{\bal}{\begin{align}}
\newcommand{\eal}{\end{align}}

\newcommand{\bcen}{\begin{center}}
\newcommand{\ecen}{\end{center}}

\newcommand{\tsp}{\textsuperscript}
\newcommand{\tsb}{\textsubscript}

\begin{document}

\title{Range-separated multideterminant density-functional theory with a short-range correlation functional of the on-top pair density}

\author{Anthony Fert\'e}
\author{Emmanuel Giner}\email{emmanuel.giner@lct.jussieu.fr}
\author{Julien Toulouse}\email{toulouse@lct.jussieu.fr}
\affiliation{
 Laboratoire de Chimie Th\'eorique (LCT), Sorbonne Universit\'e and CNRS, F-75005 Paris, France
}

\date{January 21, 2019}

\begin{abstract}
We introduce an approximation to the short-range correlation energy functional with multideterminantal reference involved in a variant of range-separated density-functional theory. This approximation is a local functional of the density, the density gradient, and the on-top pair density, which locally interpolates between the standard Perdew-Burke-Ernzerhof correlation functional at vanishing range-separation parameter and the known exact asymptotic expansion at large range-separation parameter. When combined with (selected) configuration-interaction calculations for the long-range wave function, this approximation gives accurate dissociation energy curves of the H$_2$, Li$_2$, and Be$_2$ molecules, and thus appears as a promising way to accurately account for static correlation in range-separated density-functional theory.
\end{abstract}

\maketitle

\section{Introduction}

Range-separated density-functional theory (RS-DFT) (see, e.g., Refs.~\onlinecite{Sav-INC-96,TouColSav-PRA-04}) is an alternative to Kohn-Sham density-functional theory (KS-DFT)~\cite{KohSha-PR-65} for electronic-structure calculations of atoms, molecules, and solids. It consists in rigorously combining a wave-function-type calculation for the long-range part of the Coulomb electron-electron interaction with a density functional for the complementary short-range part of the interaction. This permits to describe long-range electron correlation accurately and short-range electron correlation compactly with a fast basis-set convergence~\cite{FraMusLupTou-JCP-15}. In particular, it has been shown that explicit static correlation effects can be effectively taken into account in the long-range part of the calculation by using methods such as configuration interaction (CI)~\cite{LeiStoWerSav-CPL-97,PolSavLeiSto-JCP-02,Cas-JCP-18}, multiconfiguration self-consistent field~\cite{FroTouJen-JCP-07,FroReaWahWahJen-JCP-09,HedTouJen-JCP-18}, multireference perturbation theory~\cite{FroCimJen-PRA-10}, density-matrix functional theory~\cite{Per-PRA-10,RohTouPer-PRA-10,RohPer-JCP-11}, density-matrix renormalization group~\cite{HedKneKieJenRei-JCP-15}, or pair coupled-cluster doubles~\cite{GarBulHenScu-PCCP-15}.

A major limitation to the accuracy of RS-DFT are the semilocal density-functional approximations used for the short-range exchange-correlation energy~\cite{TouSavFla-IJQC-04,TouColSav-PRA-04,TouColSav-JCP-05,GolWerSto-PCCP-05,PazMorGorBac-PRB-06,GolWerStoLeiGorSav-CP-06}, which still suffer from self-interaction (or fractional-charge) errors and static-correlation (or fractional-spin) errors~\cite{MusTou-MP-17}. An attractive remedy to this problem is to calculate exactly a large portion of the short-range exchange-correlation energy using the multideterminant wave function naturally available in RS-DFT, leaving only a residual short-range correlation energy functional to be approximated~\cite{Tou-THESIS-05,TouGorSav-TCA-05}. In particular, this permits to drastically reduce self-interaction errors since the exchange energy is now calculated with a wave function and not with an approximate exchange density functional. This strategy has been pursued in various RS-DFT approaches~\cite{StoTeaTouHelFro-JCP-13,CorStoJenFro-PRA-13,CorFro-IJQC-14,RebTouTeaHelSav-MP-15,SenHedAlaKneFro-MP-16,AlaKneFro-PRA-16,AlaDeuKneFro-JCP-17,RebTeaHelSavTou-MP-18}. However, only a local-density approximation (LDA) for this short-range correlation energy functional with multideterminantal reference was available so far~\cite{TouGorSav-TCA-05,PazMorGorBac-PRB-06}, which tends to substantially overcorrelate.

In the present work, we develop an approximation for this short-range multideterminant correlation energy functional which uses the density, the density gradient, and the on-top pair density. The development of correlation functionals depending on the on-top pair density extracted from a multideterminant wave function has started long ago~\cite{ColSal-TCA-75,ColSal-TCA-79} and is still an active area of research (see, e.g., Ref.~\onlinecite{HolPeg-JCP-18}). An important motivation for using the on-top pair density of a multideterminant wave function is that it clearly contains information about bond dissociation (see, e.g., Ref.~\onlinecite{CarTruGag-JPCA-17}), without having to artificially break spin symmetry. In practice, most of the works in this domain introduce the on-top pair density via effective spin densities which are fed into standard spin-dependent exchange and/or correlation density functionals~\cite{MosSan-PRA-91,BecSavSto-TCA-95,MieStoSav-MP-97,TakYamYam-CPL-02,TakYamYam-IJQC-04,GraCre-MP-05,TsuScuSav-JCP-10,LimCarLuoMaOlsTruGag-JCTC-14,GarBulHenScu-JCP-15,GarBulHenScu-PCCP-15,CarTruGag-JCTC-15,GagTruLiCarHoyBa-ACR-17}. This is justified by the alternative interpretation of spin-density-functional theory~\cite{PerSavBur-PRA-95,PerErnBurSav-IJQC-97}, in which the spin densities are viewed as mere intermediate quantities for reproducing the total density and the on-top pair density. Here, instead of using effective spin densities, we introduce the dependence on the on-top pair density by exploiting the known exact asymptotic behavior of the short-range multideterminant correlation energy functional in the limit where the electron-electron interaction is very short ranged~\cite{TouGorSav-TCA-05,PazMorGorBac-PRB-06}.

The paper is organized as follows. In Sec.~\ref{theory}, we briefly review RS-DFT, including the approach involving the short-range correlation functional with multideterminant reference, and develop a new approximation for this functional. After giving computational details in Sec.~\ref{Computational details}, in particular on the selected CI method that we use for the long-range wave function, we discuss in Sec.~\ref{results} the results concerning the convergence with respect to the number of determinants on the Ne atom and the Be$_2$ molecule, the dependence on the range-separation parameter on the He and C atoms and the H$_2$ molecule near dissociation, and the dissociation energy curves of the H$_2$, Li$_2$, and Be$_2$ molecules. Finally, Sec.~\ref{conclusion} summarizes our conclusions.

\section{Theory}
\label{theory}
\subsection{Range-separated density-functional theory}

The exact ground-state energy of a $N$-electron system with nuclei-electron potential $v_\mathrm{ne}(\textbf{r})$ can be expressed by the following minimization over $N$-representable densities $n$~\cite{Lev-PNAS-79,Lie-IJQC-83}
\begin{equation}
E_0 = \min_n \left\{ \mathcal{F}[n] + \int v_\mathrm{ne}(\textbf{r}) n(\textbf{r}) \mathrm{d} \textbf{r} \right\},
\label{E0}
\end{equation}
with the standard constrained-search universal density functional
\begin{equation}
\mathcal{F}[n] = \min_{\Psi\rightarrow n} \langle\Psi|\hat{T}+\hat{W}_\mathrm{ee}|\Psi \rangle,
\label{Fn}
\end{equation}
where $\hat{T}$ and $\hat{W}_\mathrm{ee}$ are the kinetic-energy and Coulomb electron-electron interaction operators, respectively. The minimizing multideterminant wave function in Eq.~\eqref{Fn} will be denoted by $\Psi[n]$. 

In RS-DFT, the universal density functional is decomposed as~\cite{Sav-INC-96,TouColSav-PRA-04}
\begin{equation}
\mathcal{F}[n] = \mathcal{F}^{\mathrm{lr},\mu}[n] + \bar{E}_{\mathrm{Hxc}}^{\mathrm{sr,}\mu}[n],
\label{Fdecomp}
\end{equation}
where $\mathcal{F}^{\mathrm{lr},\mu}[n]$ is a long-range (lr) universal density functional
\beq
\label{lr_univ_fonc}
\mathcal{F}^{\mathrm{lr},\mu}[n]= \min_{\Psi\rightarrow n} \langle\Psi|\hat{T}+\hat{W}_\mathrm{ee}^{\mathrm{lr},\mu}|\Psi \rangle,
\eeq
and $\bar{E}_{\mathrm{Hxc}}^{\,\mathrm{sr,}\mu}[n]$ is the complementary short-range (sr) Hartree-exchange-correlation (Hxc) density functional. In Eq.~(\ref{lr_univ_fonc}), $\hat{W}_\mathrm{ee}^{\mathrm{lr}}$ is the long-range electron-electron interaction defined as
\begin{equation}
\hat{W}_\mathrm{ee}^{\mathrm{lr},\mu} = \frac{1}{2} \iint w_{\mathrm{ee}}^{\mathrm{lr},\mu}(r_{12}) \hat{n}_2(\textbf{r}_1,\textbf{r}_2)
\mathrm{d} \textbf{r}_1 \mathrm{d} \textbf{r}_2,
\end{equation}
with the error-function potential $w_{\mathrm{ee}}^{\mathrm{lr},\mu}(r_{12})=\mathrm{erf}(\mu\, r_{12} )/r_{12}$ (expressed with the interelectronic distance $r_{12} = ||\textbf{r}_1-\textbf{r}_2||$) and the pair-density operator $\hat{n}_2(\textbf{r}_1,\textbf{r}_2)=\hat{n}(\textbf{r}_1) \hat{n}(\textbf{r}_2) - \delta(\textbf{r}_1-\textbf{r}_2) \hat{n}(\textbf{r}_1)$ where $\hat{n}(\textbf{r})$ is the density operator. The range-separation parameter $\mu$ corresponds to an inverse distance controlling the range of the separation. For a given density, we will denote by $\Psi^\mu[n]$ the minimizing multideterminant wave function in Eq.~\eqref{lr_univ_fonc}. Inserting the decomposition of Eq.~\eqref{Fdecomp} into Eq.~\eqref{E0}, and recomposing the two-step minimization into a single one, leads to the following expression for the exact ground-state electronic energy
\begin{eqnarray}
\label{min_rsdft}
E_0= \min_{\Psi} \Big\{ \langle\Psi|\hat{T}+\hat{W}_\mathrm{{ee}}^{\mathrm{lr},\mu}+\hat{V}_{\mathrm{ne}}|\Psi\rangle + \bar{E}^{\mathrm{sr},\mu}_{\mathrm{Hxc}}[n_\Psi]\Big\},
\nonumber\\
\end{eqnarray}
where the minimization is done over normalized $N$-electron multideterminant wave functions, $\hat{V}_{\mathrm{ne}} = \int v_{\mathrm{ne}} (\textbf{r}) \hat{n}(\textbf{r}) \mathrm{d} \textbf{r}$, and $n_\Psi$ refers to the density of $\Psi$, i.e. $n_\Psi(\textbf{r})=\langle\Psi|\hat{n}(\textbf{r})|\Psi\rangle$. The minimizing multideterminant wave function $\Psi^\mu$ in Eq.~\eqref{min_rsdft} can be determined by the self-consistent eigenvalue equation
\beq
\label{rs-dft-eigen-equation}
\hat{H}^\mu[n_{\Psi^{\mu}}] \Ket{\Psi^{\mu}}= \mathcal{E}^{\mu} \Ket{\Psi^{\mu}},
\eeq
with the long-range interacting Hamiltonian
\beq
\label{H_mu}
\hat{H}^\mu[n_{\Psi^{\mu}}] = \hat{T}+\hat{W}_{\mathrm{ee}}^{\mathrm{lr},\mu}+\hat{V}_{\mathrm{ne}}+ \hat{\bar{V}}_{\mathrm{Hxc}}^{\mathrm{sr},\mu}[n_{\Psi^{\mu}}],
\eeq
where $\hat{\bar{V}}_{\mathrm{Hxc}}^{\mathrm{sr},\mu}[n]=\int \delta \bar{E}^{\mathrm{sr},\mu}_{\mathrm{Hxc}}[n]/\delta n(\textbf{r}) \, \hat{n}(\textbf{r}) \mathrm{d} \textbf{r}$ is the complementary short-range Hartree-exchange-correlation potential operator. Note that $\Psi^{\mu}$ is not the exact physical ground-state wave function but only an effective wave function. However, its density $n_{\Psi^{\mu}}$ is the exact physical ground-state density. Once $\Psi^{\mu}$ has been calculated, the exact electronic ground-state energy is obtained by
\beq
\label{E-rsdft}
E_0=  \braket{\Psi^{\mu}|\hat{T}+\hat{W}_\mathrm{{ee}}^{\mathrm{lr},\mu}+\hat{V}_{\mathrm{ne}}|\Psi^{\mu}}+\bar{E}^{\mathrm{sr},\mu}_{\mathrm{Hxc}}[n_{\Psi^\mu}].
\eeq
Note that, for $\mu=0$, the long-range interaction vanishes, $w_{\mathrm{ee}}^{\mathrm{lr},\mu=0}(r_{12}) = 0$, and thus RS-DFT reduces to standard KS-DFT. For $\mu\to\infty$, the long-range interaction becomes the standard Coulomb interaction, $w_{\mathrm{ee}}^{\mathrm{lr},\mu\to\infty}(r_{12}) = 1/r_{12}$, and thus RS-DFT reduces to standard wave-function theory (WFT).

In principle, Eq.~\eqref{rs-dft-eigen-equation} should be solved at the full-configuration-interaction (FCI) level in a complete one-electron basis set. In practice, however, for typical values of the range-separation parameter used (around $\mu=0.5$ bohr$^{-1}$)~\cite{GerAng-CPL-05a,FroTouJen-JCP-07}, $\hat{H}^\mu[n_{\Psi^{\mu}}]$ contains only a non-diverging soft long-range electron-electron interaction, implying that the wave function $\Psi^\mu$ does not have an electron-electron cusp~\cite{GorSav-PRA-06} and has a fast convergence with respect to the number of determinants or with respect to the size of the one-electron basis~\cite{FraMusLupTou-JCP-15}. One can then accurately solve Eq.~\eqref{rs-dft-eigen-equation} using efficient truncated or selected CI approaches, such as the configuration interaction perturbatively selected iteratively (CIPSI) method~{\cite{bender,malrieu_cipsi,buenker1,buenker-book,three_class_CIPSI,Rubio198698,cimiraglia_cipsi,harrison,cele_cipsi_zeroth_order,Angeli2000472,cipsi_property,cucl2_1,cucl2_cele}} (see Sec.~\ref{Computational details}), with relatively small basis sets. The resulting compact wave function $\Psi^\mu$ will accurately include the long-range electron correlation effects.

As regards the short-range density functional, it is usually decomposed into three contributions
\begin{eqnarray}
\bar{E}^{\mathrm{sr},\mu}_{\mathrm{Hxc}}[n] = E^{\mathrm{sr},\mu}_{\mathrm{H}}[n] + E^{\mathrm{sr},\mu}_{\mathrm{x}}[n] + \bar{E}^{\mathrm{sr},\mu}_{\mathrm{c}}[n],
\end{eqnarray}
where $E^{\mathrm{sr},\mu}_{\mathrm{H}}[n]$ is the short-range Hartree energy functional
\begin{eqnarray}
E^{\mathrm{sr},\mu}_{\mathrm{H}}[n] = \frac{1}{2} \iint w_{\mathrm{ee}}^{\mathrm{sr},\mu}(r_{12}) n(\textbf{r}_1) n(\textbf{r}_2)
\mathrm{d} \textbf{r}_1 \mathrm{d} \textbf{r}_2,
\end{eqnarray}
with the short-range electron-electron interaction $w_{\mathrm{ee}}^{\mathrm{sr},\mu}(r_{12})=1/r_{12}- w_{\mathrm{ee}}^{\mathrm{lr},\mu}(r_{12})$, and $E^{\mathrm{sr},\mu}_{\mathrm{x}}[n]$ and $\bar{E}_\mathrm{c}^{\mathrm{sr,}\mu}[n]$ are the short-range exchange and correlation energy functionals
\begin{eqnarray}
E^{\mathrm{sr},\mu}_{\mathrm{x}}[n] = \braket{\Phi^{\textsc{ks}}[n]|\hat{W}^{\mathrm{sr},\mu}_{\mathrm{ee}}|\Phi^{\textsc{ks}}[n]} - E^{\mathrm{sr},\mu}_{\mathrm{H}}[n],
\end{eqnarray}
\beq
\bar{E}_\mathrm{c}^{\mathrm{sr},\mu}[n]=\bar{E}_{\mathrm{Hxc}}^{\mathrm{sr,}\mu}[n]-\braket{\Phi^{\textsc{ks}}[n]|\hat{W}^{\mathrm{sr},\mu}_{\mathrm{ee}}|\Phi^{\textsc{ks}}[n]},
\label{Ecsr}
\eeq
defined with the Kohn-Sham (KS) single-determinant wave function $\Phi^{\textsc{ks}}[n]=\Psi^{\mu=0}[n]$ and the short-range electron-electron interaction operator
\begin{equation}
\hat{W}_\mathrm{ee}^{\mathrm{sr},\mu} = \frac{1}{2} \iint w_{\mathrm{ee}}^{\mathrm{sr},\mu}(r_{12}) \hat{n}_2(\textbf{r}_1,\textbf{r}_2)
\mathrm{d} \textbf{r}_1 \mathrm{d} \textbf{r}_2.
\end{equation}
Whereas $E^{\mathrm{sr},\mu}_{\mathrm{H}}[n]$ is calculated exactly, approximations need to be used for $E^{\mathrm{sr},\mu}_{\mathrm{x}}[n]$ and $\bar{E}_\mathrm{c}^{\mathrm{sr,}\mu}[n]$. In this work, we use the short-range version of the Perdew-Burke-Ernzerhof (PBE)~\cite{PerBurErn-PRL-96} exchange and correlation functionals of Ref.~\onlinecite{GolWerStoLeiGorSav-CP-06} (see also Refs.~\onlinecite{TouColSav-JCP-05,GolWerSto-PCCP-05}) which takes the form
\begin{eqnarray}
\bar{E}^{\mathrm{sr},\mu,\textsc{pbe}}_{\mathrm{x/c}}[n] = \int  \bar{e}_\mathrm{{x/c}}^\mathrm{sr,\mu,\textsc{pbe}}(n(\textbf{r}),\nabla n(\textbf{r})) \, \mathrm{d}\textbf{r}.
\end{eqnarray}
It has been shown that such semi-local density-functional approximations become more accurate as the range of the electron-electron interaction is reduced~\cite{TouColSav-PRA-04}. Nevertheless, for the values of the range-separation parameter commonly used, the short-range PBE exchange and correlation density functionals still contain substantial self-interaction and static-correlation errors~\cite{MusTou-MP-17}.

\subsection{Short-range correlation energy functional with multideterminant reference}

The definition of the short-range correlation energy functional in Eq.~\eqref{Ecsr} is based on the KS single-determinant wave function $\Phi^{\textsc{ks}}[n]$. In RS-DFT, it is in fact more natural to define another short-range correlation energy functional based on the multideterminant (md) wave function $\Psi^\mu[n]$~\cite{Tou-THESIS-05,TouGorSav-TCA-05}
\beq
\label{ecmd_def}
\bar{E}_\mathrm{c,md}^{\mathrm{sr},\mu}[n]=\bar{E}_{\mathrm{Hxc}}^{\mathrm{sr,}\mu}[n]-\braket{\Psi^{\mu}[n]|\hat{W}^{\mathrm{sr},\mu}_{\mathrm{ee}}|\Psi^{\mu}[n]}.
\eeq
In lieu of the standard expression of the ground-state energy in the context of RS-DFT using only the long-range electron-electron interaction in the expectation value over the wave function $\Psi^\mu$ as described by Eq.~\eqref{E-rsdft}, we can now easily include the full-range interaction in the expectation value by writing the exact ground-state electronic energy as
\beq
\label{E-ecmd}
E_0  = \braket {\Psi^{\mu}|\hat{H}|\Psi^{\mu}}+{{\bar{E}_\mathrm{{c,md}}^{\mathrm{sr,}{\mu}}[n_{\Psi^{\mu}}]}},
\eeq
where $\hat{H}=\hat{T}+\hat{W}_\mathrm{{ee}}+\hat{V}_{\mathrm{ne}}$ is the complete electronic Hamiltonian. This allows one to extract as much information as possible from the wave function $\Psi^\mu$ by calculating exactly the short-range Hartree and ``exchange'' energies related to it, i.e. the term $\braket{\Psi^{\mu}|\hat{W}^{\mathrm{sr},\mu}_{\mathrm{ee}}|\Psi^{\mu}}$. Since the wave function $\Psi^{\mu}$ is obtained without considering the short-range component of the electron-electron interaction, some short-range correlation is still missing in $\braket {\Psi^{\mu}|\hat{H}|\Psi^{\mu}}$ and must be recovered by the complementary multideterminant short-range correlation energy functional ${{\bar{E}_\mathrm{{c,md}}^{\mathrm{sr,}{\mu}}[n]}}$. Obviously, in practice, this functional must be approximated, but calculating the energy via Eq.~\eqref{E-ecmd} instead of Eq.~\eqref{E-rsdft} reduces the demand put on density-functional approximations. In particular, since in Eq.~\eqref{E-ecmd} the whole exchange energy is calculated with a wave function and not with an approximate exchange density functional, we expect to eliminate most of the self-interaction error. We note that, contrary to the expression in Eq.~\eqref{E-rsdft}, the energy expression in Eq.~\eqref{E-ecmd} is not variational with respect to $\Psi^\mu$. Even though it is possible to formulate a self-consistent version of Eq.~\eqref{E-ecmd} via a multideterminant extension of the optimized-effective-potential (OEP) approach~\cite{TouGorSav-TCA-05,StoTeaTouHelFro-JCP-13}, we do not consider this possibility in this work.

In order to construct an approximation for ${{\bar{E}_\mathrm{{c,md}}^{\mathrm{sr,}{\mu}}[n]}}$, we now study two exact conditions on this functional. For this, it is convenient to express the functional $\bar{E}_\mathrm{c,md}^{\mathrm{sr},\mu}[n]$ in terms of the original functional $\bar{E}_\mathrm{c}^{\mathrm{sr},\mu}[n]$, using Eqs.~\eqref{Ecsr} and~\eqref{ecmd_def},
\beq
\label{ecmd}
\bar{E}_\mathrm{c,md}^{\mathrm{sr},\mu}[n]=\bar{E}^{\mathrm{sr},\mu}_{\mathrm{c}}[n] + \Delta^{\mathrm{lr}\textsc{-}\mathrm{sr},\mu}[n],
\eeq
where $\Delta^{\mathrm{lr}\textsc{-}\mathrm{sr},\mu}[n]$ is a mixed long-range/short-range quantity
\begin{align}
\label{delta_ecmd}
\Delta^{\mathrm{lr}\textsc{-}\mathrm{sr},\mu}[n]= & \braket{\Phi^{\textsc{ks}}[n]|\hat{W}_{\mathrm{ee}}^{\mathrm{sr},\mu}| \Phi^{\textsc{ks}}[n]}
\nonumber\\
&\;\;\;\;\;- \braket {\Psi^{\mu}[n]|\hat{W}_{\mathrm{ee}}^{\mathrm{sr},\mu}|\Psi^{\mu}[n]}.
\end{align}
We expect for most systems that $\Delta^{\mathrm{lr}\textsc{-}\mathrm{sr},\mu} \geq 0$, i.e. $|\bar{E}_\mathrm{c,md}^{\mathrm{sr},\mu}| \leq |\bar{E}^{\mathrm{sr},\mu}_{\mathrm{c}}|$.

The first condition is for $\mu=0$. In this case, since the RS-DFT wave function reduces to the KS wave function, $\Psi^{\mu=0}[n]=\Phi^{\textsc{ks}}[n]$, the short-range multideterminant correlation functional reduces to the usual KS correlation functional
\beq
\bar{E}_\mathrm{{c,md}}^\mathrm{{\,sr},\mu=0}[n] =\bar{E}^{\,\mathrm{sr},\mu=0}_{\mathrm{c}}[n] = E_\mathrm{c}^{\textsc{ks}}[n].
\label{Ecmdmu0}
\eeq

The second condition is for $\mu\to\infty$. In this limit, the asymptotic expansion of $\bar{E}^{\mathrm{sr},\mu}_{\mathrm{c}}[n]$ is known to be~\cite{TouColSav-PRA-04,GorSav-PRA-06}
\begin{eqnarray}
\label{ec_mu_inf}
\bar{E}^{\mathrm{sr},\mu}_{\mathrm{c}}[n] &=&  \frac{\pi}{2\mu^2}\int \!\! n_{2,\mathrm{c}}(\textbf{r},\textbf{r})\mathrm{d}\textbf{r}
\nonumber\\
&&+  \frac{2\sqrt{2\pi}}{3\mu^3} \int \!\! n_{2}(\textbf{r},\textbf{r})\mathrm{d}\textbf{r} + O\left( \frac{1}{\mu^4} \right), \;\;\;\;
\end{eqnarray}
where $n_{2}(\textbf{r},\textbf{r})=\braket{\Psi[n]|\hat{n}_2(\textbf{r},\textbf{r})|\Psi[n]}$ is the Coulombic on-top pair density (i.e., the on-top pair density associated with the full-range wave function $\Psi[n]$) and $n_{2,\mathrm{c}}(\textbf{r},\textbf{r})=n_{2}(\textbf{r},\textbf{r})-n_{2,\textsc{ks}}(\textbf{r},\textbf{r})$ is its correlation contribution defined with respect to the KS on-top pair density $n_{2,\textsc{ks}}(\textbf{r},\textbf{r})=\braket{\Phi^{\textsc{ks}}[n]|\hat{n}_2(\textbf{r},\textbf{r})|\Phi^{\textsc{ks}}[n]}$. The asymptotic expansion of $\Delta^{\mathrm{lr}\textsc{-}\mathrm{sr},\mu}[n]$ for $\mu\to\infty$ can be obtained by generalizing the expansion given in the case of the homogeneous electron gas in Ref.~\onlinecite{PazMorGorBac-PRB-06}, leading to
\begin{eqnarray}
\label{delta_mu_inf}
\Delta^{\mathrm{lr}\textsc{-}\mathrm{sr},\mu}[n] = -\frac{\pi}{2\mu^2} \int \!\! n_{2,\mathrm{c}}(\textbf{r},\textbf{r})\mathrm{d}\textbf{r} \phantom{xxxxxxxxxxxx}
\nonumber\\
- \frac{2\sqrt{\pi}(2\sqrt{2}-1)}{3\mu^3} \int \!\! n_{2}(\textbf{r},\textbf{r})\mathrm{d}\textbf{r}
+O \left(\frac{1}{\mu^4}\right). \;\;
\end{eqnarray}
The terms in $1/\mu^{2}$ in Eqs.~\eqref{ec_mu_inf} and~\eqref{delta_mu_inf} cancel each other, and we get the asymptotic expansion of $\bar{E}_\mathrm{{c,md}}^\mathrm{{sr},\mu}[n]$ for $\mu\to\infty$
\beq
\label{Ec-md-inf}
\bar{E}_\mathrm{{c,md}}^\mathrm{{sr},\mu}[n] = \frac{2\sqrt{\pi}(1-\sqrt{2})}{3\mu^3} \int \!\! n_{2}(\textbf{r},\textbf{r})\mathrm{d}\textbf{r}
+O \left(\frac{1}{\mu^4}\right).
\eeq
The short-range multideterminant correlation functional $\bar{E}_\mathrm{{c,md}}^\mathrm{{sr},\mu}[n]$ goes to zero as $1/\mu^{3}$ when $\mu\to\infty$, i.e. faster that the original short-range correlation functional $\bar{E}^{\mathrm{sr},\mu}_{\mathrm{c}}[n]$ of RS-DFT. This is not a surprise since $\bar{E}_\mathrm{{c,md}}^\mathrm{{sr},\mu}[n]$ accounts for a smaller part of the correlation energy than $\bar{E}^{\mathrm{sr},\mu}_{\mathrm{c}}[n]$. We thus see that, because of the local nature of the short-range interaction for a large value of $\mu$, the on-top pair density $n_{2}(\textbf{r},\textbf{r})$ appears naturally as a key ingredient in the short-range multideterminant correlation functional $\bar{E}_\mathrm{{c,md}}^\mathrm{{sr},\mu}[n]$.

\subsection{Approximations for the short-range multideterminant correlation functional $\bar{E}_\mathrm{{c,md}}^\mathrm{{sr},\mu}[n]$}

Until now, the only approximation available for the functional $\bar{E}_\mathrm{{c,md}}^\mathrm{{sr},\mu}[n]$ was the short-range LDA (srLDA) approximation~\cite{TouGorSav-TCA-05,PazMorGorBac-PRB-06}
\beq
\bar{E}_\mathrm{{c,md}}^\mathrm{{sr},\mu,\textsc{lda}}[n]= \int \bar{e}_\mathrm{{c,md}}^\mathrm{{sr},\mu,\textsc{lda}}(n(\textbf{r})) \, \mathrm{d}\textbf{r},
\label{Ecmdlda}
\eeq
where $\bar{e}_\mathrm{{c,md}}^\mathrm{{sr},\mu,\textsc{lda}}(n)$ is the energy density extracted from the homogeneous electron gas for which a parametrization is given in Ref.~\onlinecite{PazMorGorBac-PRB-06}. Unfortunately, this srLDA approximation tends to give substantially too negative correlation energies for small values of $\mu$ (and in particular for the values commonly used, i.e. around $\mu=0.5$ bohr$^{-1}$)~\cite{TouGorSav-TCA-05,StoTeaTouHelFro-JCP-13}.

Here, we construct a new approximation for the functional $\bar{E}_\mathrm{{c,md}}^\mathrm{{sr},\mu}[n]$ based on the two exact conditions in Eqs.~\eqref{Ecmdmu0} and~\eqref{Ec-md-inf}. We propose a local interpolation between the standard PBE correlation functional at $\mu=0$ (of course, any other generalized-gradient approximation to the KS correlation functional could be used) and the leading term of the asymptotic expansion of $\bar{E}_\mathrm{{c,md}}^\mathrm{{sr},\mu}[n]$ for $\mu\to\infty$. The resulting approximation, referred to as ``srPBEontop'', is a local functional of the density, the density gradient, and the on-top pair density
\begin{align}
\bar{E}_{\mathrm{c,md}}^{{\mathrm{sr,}\mu},\textsc{pbe}\text{\tiny ontop}}[n] =   \phantom{xxxxxxxxxxxxxxx}  &
\nonumber\\
\int  \bar{e}_{\mathrm{c,md}}^{{\mathrm{sr,}\mu},\textsc{pbe}\text{\tiny ontop}}(n(\textbf{r}),\nabla n(\textbf{r}),n_2(\textbf{r},\textbf{r}))  \, &\mathrm{d}\textbf{r},
\label{EcmdsrPBEontop}
\end{align}
where the energy density is taken as 
\beq
\label{e_srPBEontop}
\bar{e}_{\mathrm{c,md}}^{{\mathrm{sr,}\mu},\textsc{pbe}\text{\tiny ontop}}(n,\nabla n,n_2) = \frac{e_\mathrm{c}^{\textsc{pbe}}(n,\nabla n)}{1+ \beta(n,\nabla n,n_2) \mu^3},
\eeq
which reduces to the standard PBE correlation energy density $e_\mathrm{c}^{\textsc{pbe}}(n,\nabla n)$ for $\mu=0$. In order to recover the correct large-$\mu$ behavior in Eq.~\eqref{Ec-md-inf}, $\beta(n,\nabla n,n_2)$ is taken as
\beq
\label{beta_srPBEontop}
\beta(n,\nabla n,n_2) = \frac{3\,e_\mathrm{c}^{\textsc{pbe}}(n,\nabla n)}{2\sqrt{\pi}(1-\sqrt{2})n_2}.
\eeq

However, there is one difficulty with using the approximation in Eq.~\eqref{EcmdsrPBEontop}: the Coulombic on-top pair density $n_2(\textbf{r},\textbf{r})$ is not available in RS-DFT. Instead, what is available is the on-top pair density of the wave function $\Psi^\mu$ obtained with a long-range electron-electron interaction: $n_2^\mu(\textbf{r},\textbf{r}) = \braket{\Psi^\mu|\hat{n}_2(\textbf{r},\textbf{r})|\Psi^\mu}$. Fortunately, the Coulombic on-top pair density $n_2(\textbf{r},\textbf{r})$ can be extrapolated from the long-range on-top pair density $n_2^\mu(\textbf{r},\textbf{r})$, as shown in Ref.~\onlinecite{GorSav-PRA-06}. 
The extrapolation is based on the asymptotic expansion of $n_2^\mu(\textbf{r},\textbf{r})$ for $\mu\to\infty$~\cite{GorSav-PRA-06}
\beq
n_2^\mu(\textbf{r},\textbf{r})=n_2(\textbf{r},\textbf{r})\left(1+\frac{2}{\sqrt{\pi}\mu}\right)+O\left(\frac{1}{\mu^2}\right),
\eeq
which, after inversion, gives the following estimation of the Coulombic on-top pair density
\beq
\label{correction-limite}
n_2(\textbf{r},\textbf{r})\approx n_2^\mu(\textbf{r},\textbf{r})\left(1+\frac{2}{\sqrt{\pi}\mu}\right)^{-1}.
\eeq
Obviously, the Coulombic on-top pair density $n_2(\textbf{r},\textbf{r})$ is smaller than the long-range one $n_2^\mu(\textbf{r},\textbf{r})$. Since the latter is obtained with a reduced electron-electron repulsion, the probability of finding two electrons at the same point of space is larger. Note that, in the limit $\mu=0$, the extrapolation formula in Eq.~\eqref{correction-limite} just unphysically gives $n_2(\textbf{r},\textbf{r})=0$ for all systems. However, this is not a problem since in the srPBEontop functional of Eq.~\eqref{EcmdsrPBEontop} the on-top pair density $n_2(\textbf{r},\textbf{r})$ has an effect only for not too small values of $\mu$. In Ref.~\onlinecite{GorSav-PRA-06}, another extrapolation method based on the pair-distribution function of the homogeneous electron gas was also proposed. We do not consider this latter extrapolation in the present work since we have found that the simple one in Eq.~\eqref{correction-limite} gives satisfying results.

For one-electron systems, the on-top pair density $n_2(\textbf{r},\textbf{r})$ or $n_2^\mu(\textbf{r},\textbf{r})$ vanishes, and consequently, from Eqs.~\eqref{e_srPBEontop} and~\eqref{beta_srPBEontop}, the srPBEontop correlation energy correctly vanishes as well. In other words, the srPBEontop correlation functional is self-interaction free for one-electron systems. In many-electron systems, we expect the same behavior in spatial regions of one-electron character.

\section{Computational details}
\label{Computational details}

We have implemented the RS-DFT approach, including the short-range multideterminant correlation functionals, in the software \textsc{Quantum Package}~\cite{quantum_package}. 

In practice, we first perform calculations with the self-consistent RS-DFT approach of Eq.~\eqref{min_rsdft} using the srPBE approximation of Ref.~\onlinecite{GolWerStoLeiGorSav-CP-06} for the short-range exchange-correlation functional $\bar{E}_\mathrm{{xc}}^{\mathrm{sr,}{\mu}}[n]$. We calculate the multideterminant wave function $\Psi^\mu$ by solving Eq.~\eqref{rs-dft-eigen-equation} at the FCI or CIPSI level (see below) using Hartree-Fock orbitals. After a FCI or CIPSI calculation, the density $n_{\Psi^{\mu}}$ entering the short-range Hartree-exchange-correlation potential $\hat{\bar{V}}_{\mathrm{Hxc}}^{\mathrm{sr},\mu}[n_{\Psi^{\mu}}]$ in Eq.~(\ref{H_mu}) is updated and the procedure is iterated to achieve convergence with respect to the density (with an energy threshold of 10$^{-4}$ hartree). Depending on the type of long-range CI calculation used, we will refer to this method as ``lrFCI+srPBE'' or ``lrCIPSI+srPBE'', or generically as ``lrCI+srPBE''.

We then perform calculations according to Eq.~\eqref{E-ecmd} with the previously calculated wave function $\Psi^\mu$ and using either the srLDA approximation of Eq.~\eqref{Ecmdlda} or the srPBEontop approximation of Eq.~\eqref{EcmdsrPBEontop} for the short-range multideterminant correlation functional $\bar{E}_\mathrm{{c,md}}^{\mathrm{sr,}{\mu}}[n]$. We will refer to these calculations as ``CI+E\tsb{c,md}(srLDA)'' and ``CI+E\tsb{c,md}(srPBEontop)'', where again CI can stand for either FCI or CIPSI.

We now briefly describe the CIPSI method as used here. The CIPSI method~{\cite{bender,malrieu_cipsi,buenker1,buenker-book,three_class_CIPSI,Rubio198698,cimiraglia_cipsi,harrison,cele_cipsi_zeroth_order,Angeli2000472,cipsi_property,cucl2_1,cucl2_cele}} is a selected CI which allows one to perform wave-function calculations at the near FCI level by keeping only the most important Slater determinants in a given FCI space. Starting from an initial guess for the wave function, $\Ket{\Psi^{\mu,(0)}}=\sum_{\mathrm{I}\in\mathcal{R}} c^\mu_\textsc{i} \Ket{\mathrm{I}}$ where $\Ket{\mathrm{I}}$ are Slater determinants in the reference variational space $\mathcal{R}$, the importance of a given Slater determinant $\Ket{\mathrm{K}}$ outside $\mathcal{R}$ is estimated using Epstein-Nesbet multireference perturbation theory. The second-order correction on the eigenvalue associated with the reference wave function $\ \mathcal{E}^{\mu,(0)}$ arising from the Slater determinant $\Ket{\mathrm{K}}$ is given by
\beq
{\cal E}_\text{K}^{\mu,(2)}= \frac{|\braket{\Psi^{\mu,(0)}|\hat{H}^\mu|\mathrm{K}}|^2}{ \mathcal{E}^{\mu,(0)} -  \braket{\mathrm{K}|\hat{H}^\mu |\mathrm{K}}}.
\eeq
The variational space $\mathcal{R}$ is then enlarged by including the determinants associated with the largest perturbative corrections, and the procedure is iterated. In practice, the size of the variational space is doubled at each iteration until the magnitude of the total second-order Epstein-Nesbet correction on the eigenvalue, $\mathcal{E}^{\mu,(2)}=\sum_\mathrm{K} {\cal E}^{\mu,(2)}_\text{K}$, is smaller than a given threshold (10$^{-5}$ hartree or smaller). 
At a given iteration of the loop over the density $n_{\Psi^{\mu}}$ (entering the short-range potential $\hat{\bar{V}}_{\mathrm{Hxc}}^{\mathrm{sr},\mu}[n_{\Psi^{\mu}}]$), we use the wave function obtained at the previous iteration as the starting guess for the CIPSI calculation. Thus, the variational space considered at the $i$\tsp{th} iteration is included in the variational space considered at the $(i+1)^\text{th}$ iteration, $\mathcal{R}^{(i)}\subset\mathcal{R}^{(i+1)}$. Note that in order to fully couple the RS-DFT calculation with the CIPSI method we should then add a perturbative correction to the total energy in Eqs. (\ref{E-rsdft}) or (\ref{E-ecmd}). However, in the present study, since all CIPSI calculations were iterated until we obtained a very small $\mathcal{E}^{\mu,(2)}$, we can neglect this perturbative correction to the total energy in comparison to the threshold used for converging the density. 

All calculations were performed using correlation-consistent Dunning basis sets \cite{Dun-JCP-89,WooDun-JCP-95,PraWooPetDunWil-TCA-11} specified later.

\section{Results and discussion}
\label{results}

\subsection{Convergence with respect to the number of determinants}
\label{convergence}

\begin{figure*}[t]
        \centering
        \includegraphics[scale=0.6]{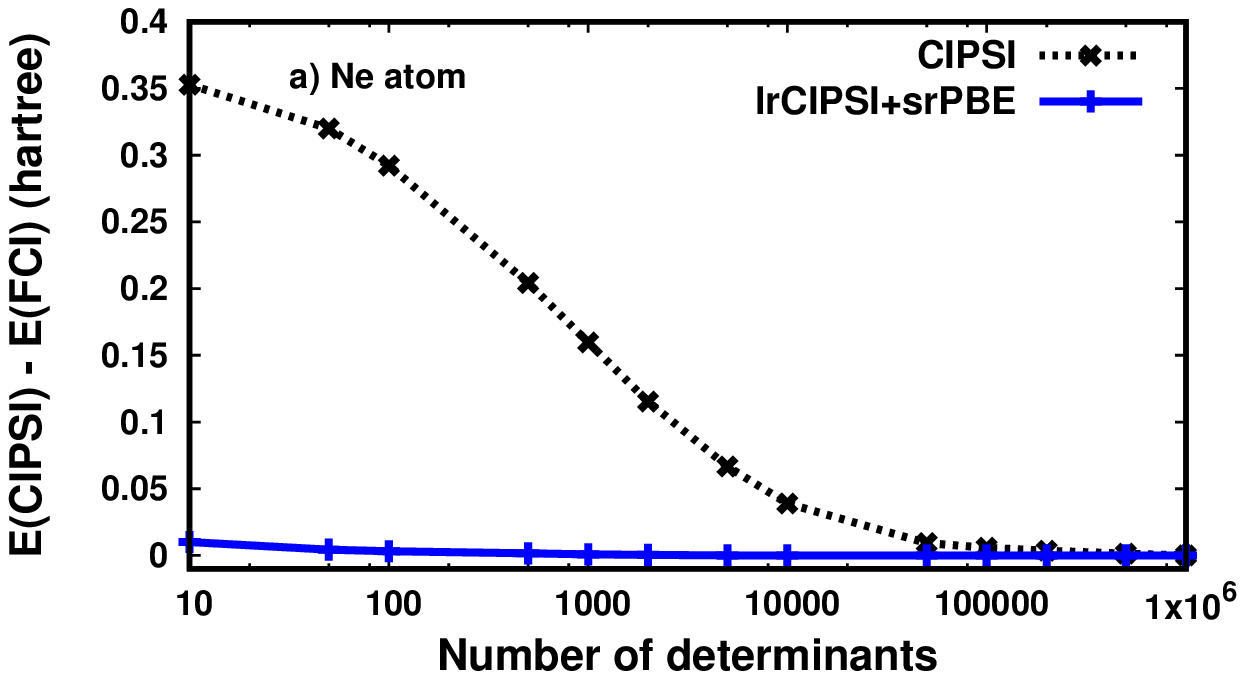}     
        \includegraphics[scale=0.6]{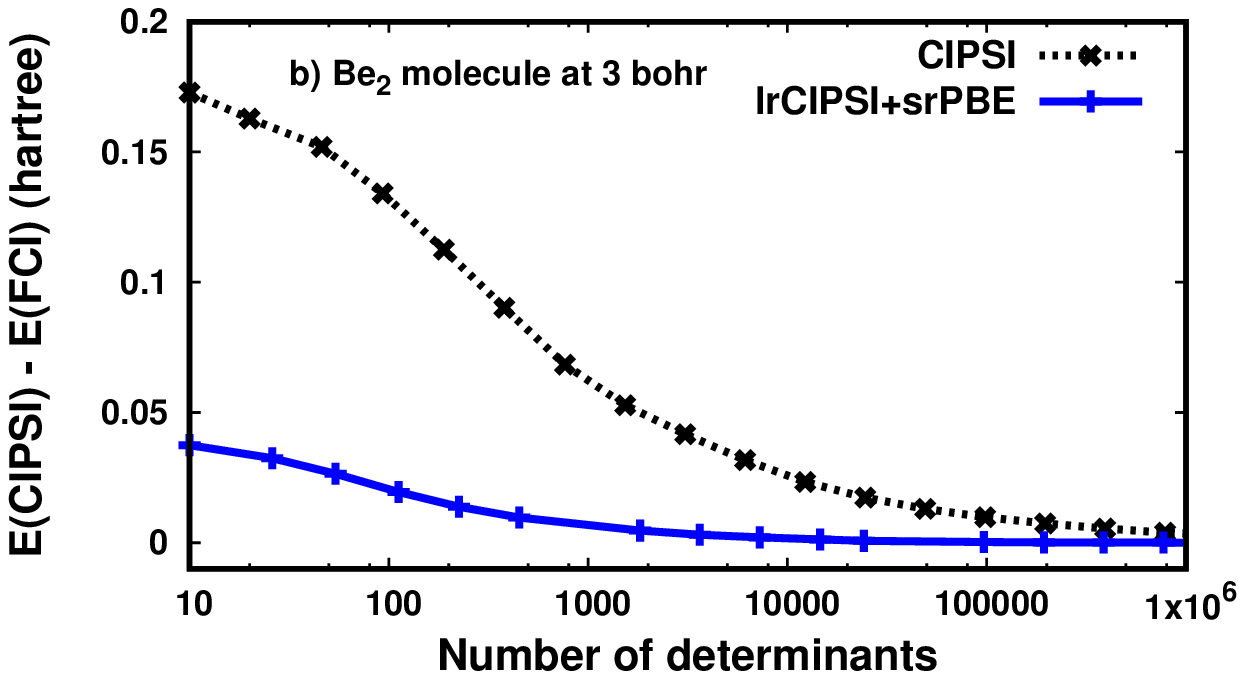}     
        \caption{Convergence of the standard CIPSI and lrCIPSI+srPBE total variational energies (measured with respect to their respective FCI limits) as a function of the number of selected determinants for a) the Ne atom with the aug-cc-pCVQZ basis set and b) the Be$_2$ molecule (internuclear distance of 3 bohr) with aug-cc-pCVTZ basis set. All electrons are correlated. The range-separation parameter used is $\mu=0.5\ \mathrm{bohr}^{-1}$.}
		\label{convergence}
\end{figure*}

We first report in Fig.~\ref{convergence} a comparison of the convergence of the standard CIPSI and lrCIPSI+srPBE total variational energies as a function of the number of selected determinants for the Ne atom and for the Be$_2$ molecule using the aug-cc-pCVQZ and aug-cc-pCVTZ basis sets, respectively, and correlating all the electrons in the CI calculations. For lrCIPSI+srPBE we use a range-separation parameter of $\mu=0.5\ \mathrm{bohr}^{-1}$. This figure clearly illustrates that the cuspless long-range wave function $\Psi^\mu$ of RS-DFT is much more compact than its Coulombic counterpart of standard WFT. Indeed, one sees that with a mere hundreds or thousands of determinants the lrCIPSI+srPBE total energy is already converged as much as the standard CIPSI total energy is with hundreds of thousands of determinants. This shows that the coupling of RS-DFT with a selected CI procedure such as the CIPSI method allows one to reduce by several orders of magnitude the dimension of the variational space of the wave function required to obtain a given accuracy.

\subsection{Total energies as a function of the range-separation parameter}

\begin{figure}[t]
        \centering
        \includegraphics[scale=0.6]{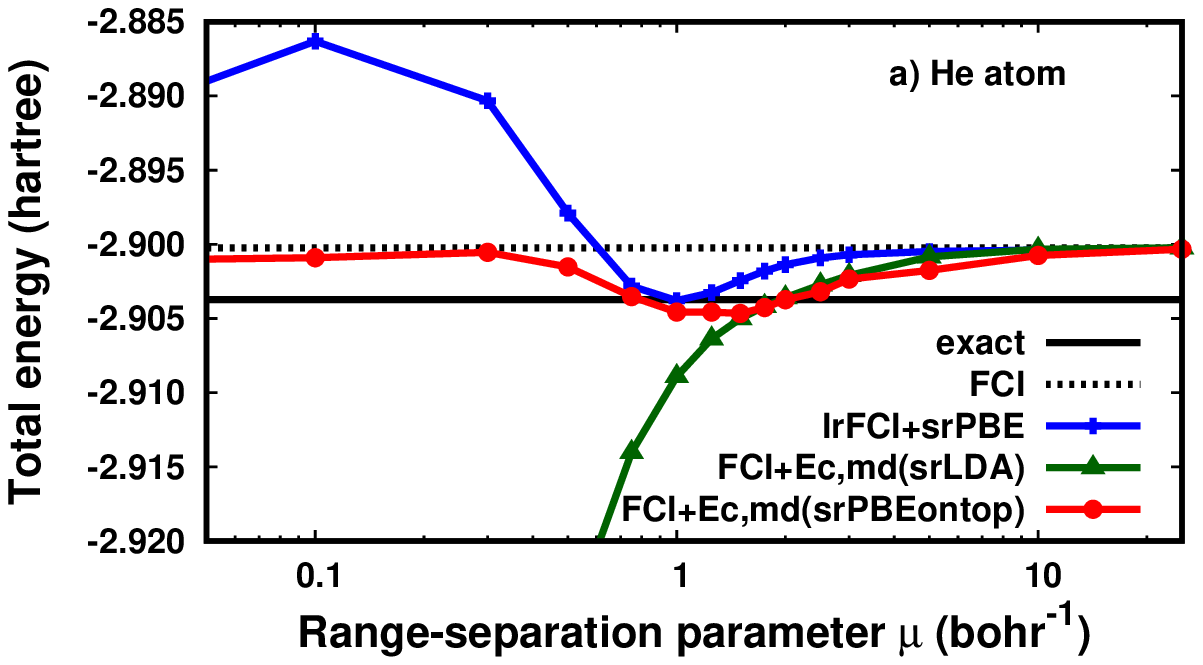}     
        \includegraphics[scale=0.6]{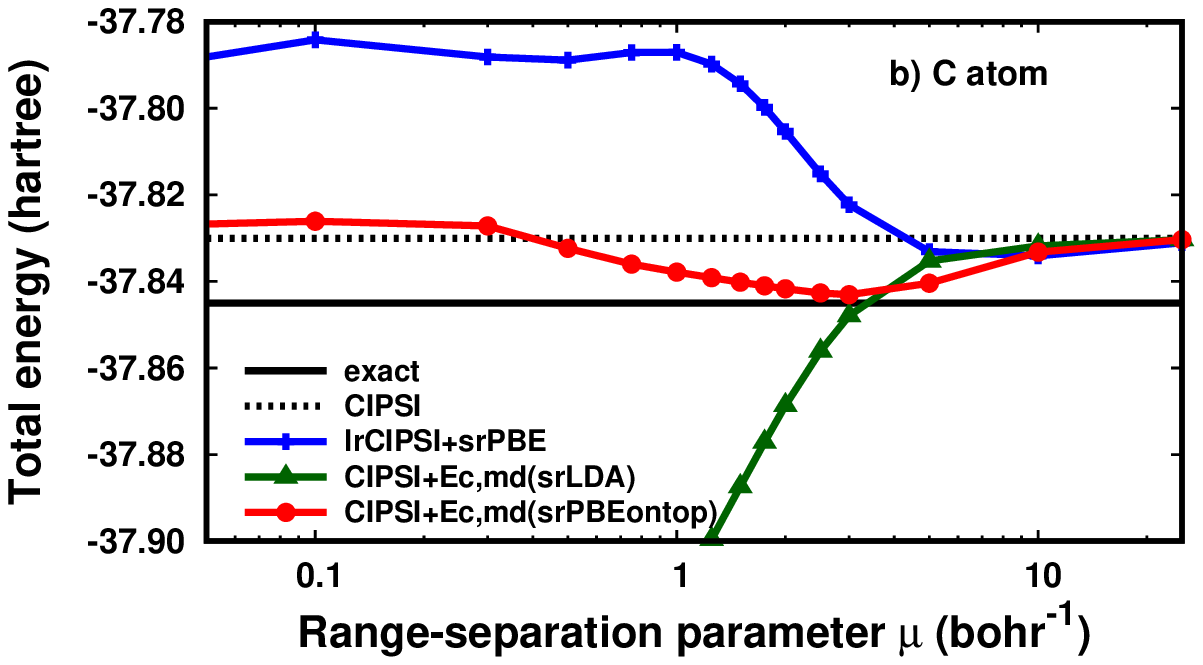}     
        \includegraphics[scale=0.6]{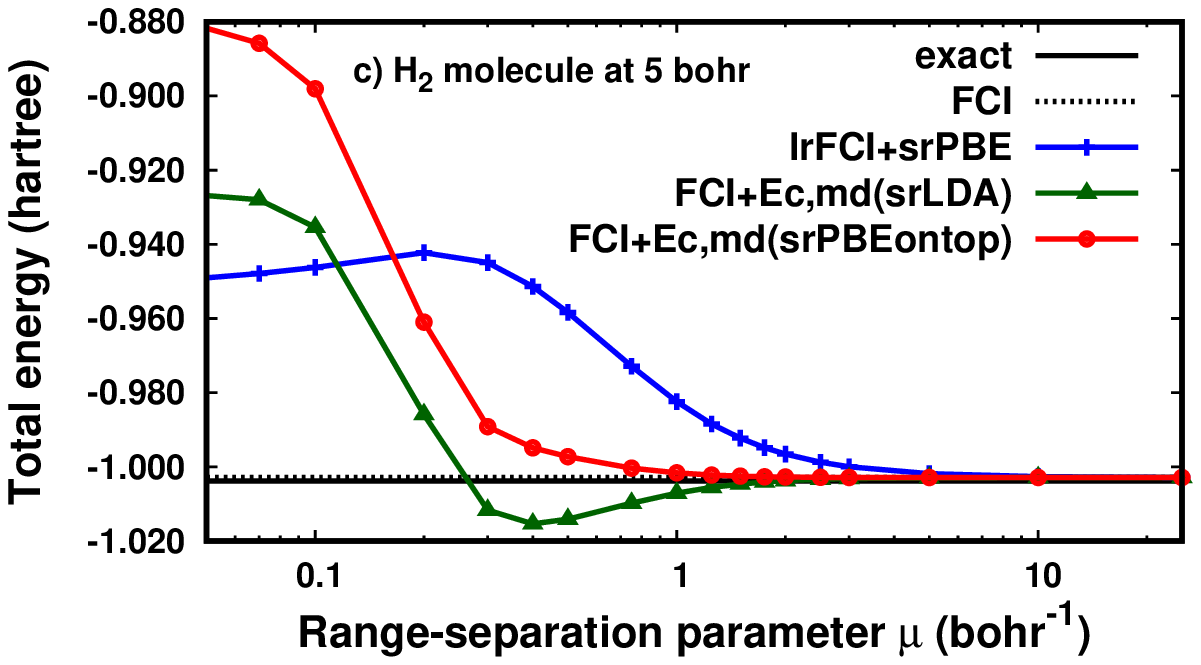}     
        \caption{Total energy of the a) He atom, b) C atom, and c) H$_2$ molecule near dissociation (internuclear distance of 5 bohr) calculated by lrCI+srPBE, CI+E\tsb{c,md}(srLDA), and CI+E\tsb{c,md}(srPBEontop) as a function of the range-separation parameter $\mu$. For He and H$_2$, the basis set used is cc-pVTZ. For C, the basis set used is cc-pCVTZ and the core excitations are allowed. For comparison, the estimated exact non-relativistic energy~\cite{DavHagChaMeiFro-PRA-91,ChaGwaDavParFro-PRA-93,LieCle-JCP-74a} and the FCI or well-converged CIPSI energy obtained with the same basis set are also reported.}
		\label{E_mu}
\end{figure}

We now discuss the accuracy of the total energy obtained with the different approximate RS-DFT schemes as a function of the range-separation parameter $\mu$. Figure~\ref{E_mu} reports the results for the He and C atoms, and for the H$_2$ molecule near dissociation as an example of a strongly correlated system. For He and H$_2$, the calculations were performed at the FCI level using the cc-pVTZ basis set. For C, the calculations were performed at the CIPSI level using the cc-pCVTZ basis set and allowing core excitations.

\subsubsection{lrCI+srPBE total energy}
\label{RSDFT intermediate mu}

As previously noted, RS-DFT reduces to KS-DFT for $\mu = 0$ and to standard WFT for $\mu \to \infty$. The behavior of the lrCI+srPBE total energy as a function of $\mu$ is in agreement with these limits. Indeed, for $\mu \to 0$, the lrCI+srPBE energy is substantially above the CI energy and goes toward the KS-DFT energy obtained with the PBE exchange-correlation functional (KS-PBE, not shown). For $\mu \to \infty$, the lrCI+srPBE energy converges asymptotically to the standard CI energy.

For an optimal intermediate value of $\mu$, which is dependent on the system, the lrCI+srPBE total energy is comparable to the full-range CI total energy, or even more accurate in the case of He at the energy minimum. We must stress, however, that since the lrCI+srPBE total energy is not necessarily an upper bound of the exact energy, the value of $\mu$ minimizing the total energy cannot generally be considered as the optimal value of $\mu$. For the cases of C and H$_2$, the lrCI+srPBE total energy is not more accurate (or only marginally) than the CI total energy, but the use of lrCI+srPBE allows one to obtain near FCI quality results with not too large a value of $\mu$ leading to a more compact wave function, as discussed in Sec.~\ref{convergence}.

We note that the optimal value of $\mu$ required to obtain an accurate total energy is substantially larger for C than for He and H$_2$. This is due to the contribution of the core spatial region of C which is associated to high densities and thus to small interelectronic distances. In order to have a part of the exchange-correlation energy of the core electrons of C treated via the long-range CI wave function, the long-range electron-electron interaction $w_{\mathrm{ee}}^{\mathrm{lr},\mu}(r_{12})$ must include the interaction between electrons at sufficiently small distances, i.e. $\mu$ must be sufficiently large.

\subsubsection{CI+E\tsb{c,md}(srLDA) total energy}

For $\mu \to \infty$, the CI+E\tsb{c,md}(srLDA) total energy converges to the standard CI total energy, as was the case for lrCI+srPBE. However, the $\mu = 0$ limit is different. Since the whole exchange energy is extracted from the wave function in the approach using the short-range multideterminantal correlation functional [Eq.~(\ref{E-ecmd})], the $\mu = 0$ limit corresponds to a KS-DFT calculation with exact exchange. More precisely, at $\mu=0$, the CI+E\tsb{c,md}(srLDA) energy reduces to $\braket{\Phi^{\textsc{pbe}}|\hat{H}|\Phi^{\textsc{pbe}}}+E_\mathrm{c}^{\textsc{lda}}[n_{\Phi^{\textsc{pbe}}}]$ where $\Phi^{\textsc{pbe}}$ is the KS single determinant obtained by solving the KS equation with the PBE exchange-correlation functional, and $E_\mathrm{c}^{\textsc{lda}}[n]$ is the standard LDA correlation functional. This explains why, for small values of $\mu$, CI+E\tsb{c,md}(srLDA) is inaccurate. For He and C, it gives far too negative total energies because the well-known overestimation (in absolute value) of the correlation energy by the LDA functional by about a factor of 2 is not compensated by an approximate exchange functional as in standard KS-DFT. For H$_2$ near dissociation, the missing static correlation effects makes the CI+E\tsb{c,md}(srLDA) total energy too high for small values of $\mu$.

Due to its very inaccurate $\mu = 0$ limit, the CI+E\tsb{c,md}(srLDA) total energies tend to have large variations with respect to $\mu$ and become reasonably accurate only for values of $\mu$ similar to those required for lrCI+srPBE. Therefore, CI+E\tsb{c,md}(srLDA) cannot be considered as an improvement over lrCI+srPBE. A better approximation must be used for the short-range multideterminantal correlation functional ${{\bar{E}_\mathrm{{c,md}}^{\mathrm{sr,}{\mu}}[n]}}$.

\subsubsection{CI+E\tsb{c,md}(srPBEontop) total energy}

As was the case for CI+E\tsb{c,md}(srLDA), the CI+E\tsb{c,md}(srPBEontop) total energy goes to the standard CI total energy for $\mu\to\infty$. For $\mu=0$, the CI+E\tsb{c,md}(srPBEontop) energy reduces to the KS-PBE energy with exact exchange, i.e. $\braket{\Phi^{\textsc{pbe}}|\hat{H}|\Phi^{\textsc{pbe}}}+E_\mathrm{c}^{\textsc{pbe}}[n_{\Phi^{\textsc{pbe}}}]$ where $E_\mathrm{c}^{\textsc{pbe}}$ is the standard PBE correlation functional. By contrast, we remind that lrCI+srPBE reduces to standard KS-PBE at $\mu=0$. One must have this in mind when comparing CI+E\tsb{c,md}(srPBEontop) and lrCI+srPBE at small $\mu$. 

For He and C, it turns out that KS-PBE with exact exchange is more accurate than standard KS-PBE, which makes CI+E\tsb{c,md}(srPBEontop) more accurate than lrCI+srPBE at small and intermediate $\mu$. Also, the fact that the PBE approximation to the KS correlation functional is a better approximation than the LDA approximation makes CI+E\tsb{c,md}(srPBEontop) give much more accurate total energies at small and intermediate $\mu$ in comparison to CI+E\tsb{c,md}(srLDA). For these weakly correlated systems, CI+E\tsb{c,md}(srPBEontop) gives quite accurate total energies over the whole range of $\mu$.

For H$_2$ near dissociation, because of the presence of static correlation effects, the total energy given by CI+E\tsb{c,md}(srPBEontop) is much too high for small values of $\mu$. In particular, at $\mu=0$, we recover the known fact that KS-PBE with exact exchange gives a larger error than KS-PBE for strongly correlated systems. However, the error rapidly decreases with $\mu$, and the CI+E\tsb{c,md}(srPBEontop) total energy converges to the accurate CI total energy significantly faster than both lrCI+srPBE and CI+E\tsb{c,md}(srLDA). This must be due to the use of the on-top pair density which imposes the correct asymptotic behavior for $\mu\to\infty$.

We thus conclude that the srPBEontop approximation to the short-range multideterminant correlation functional ${{\bar{E}_\mathrm{{c,md}}^{\mathrm{sr,}{\mu}}[n]}}$ constitutes overall a large improvement over the srLDA approximation.

\subsection{Dissociation energy curves of the H$_2$, Li$_2$, and Be$_2$ molecules}

\begin{figure*}[t]
   \begin{minipage}[c]{.46\linewidth}
      \centering
      \includegraphics[scale=0.6]{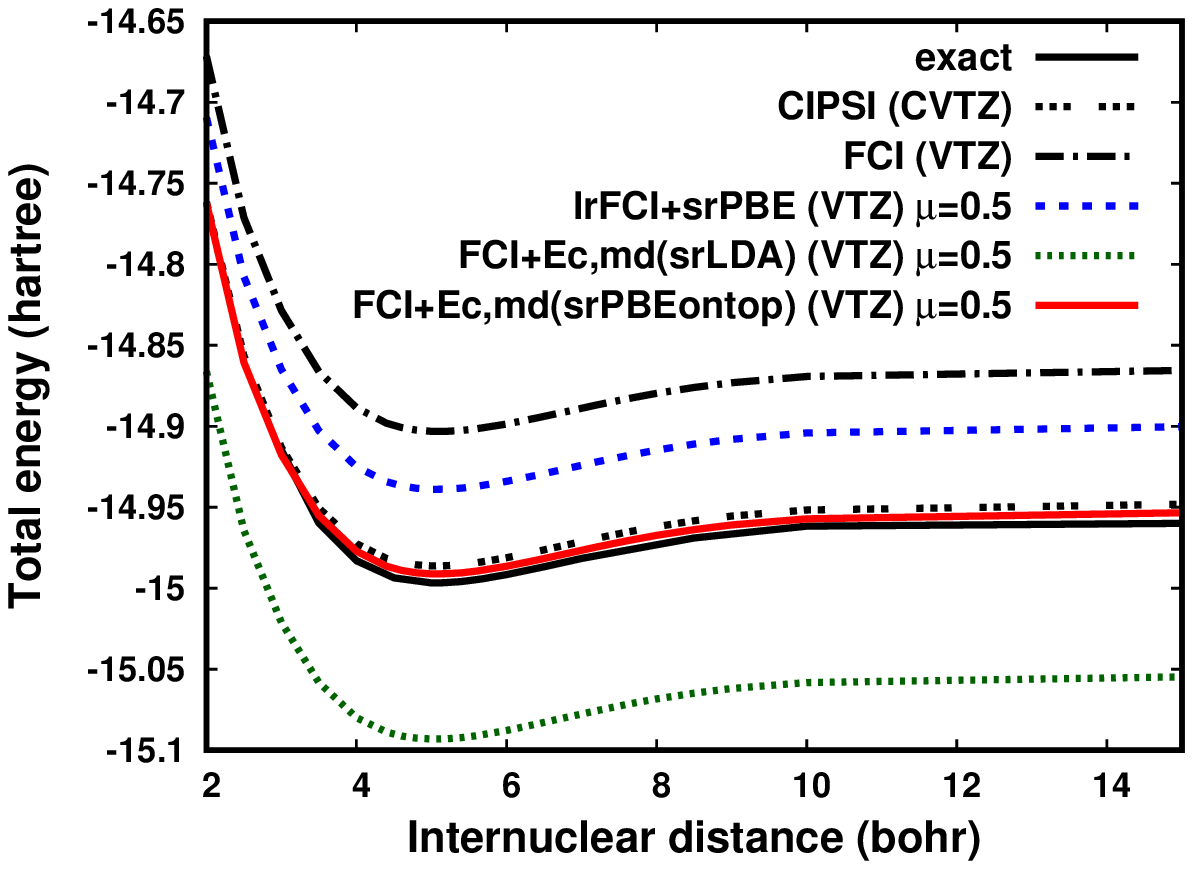}     
   \end{minipage} \hfill
   \begin{minipage}[c]{.46\linewidth}
      \centering
      \includegraphics[scale=0.6]{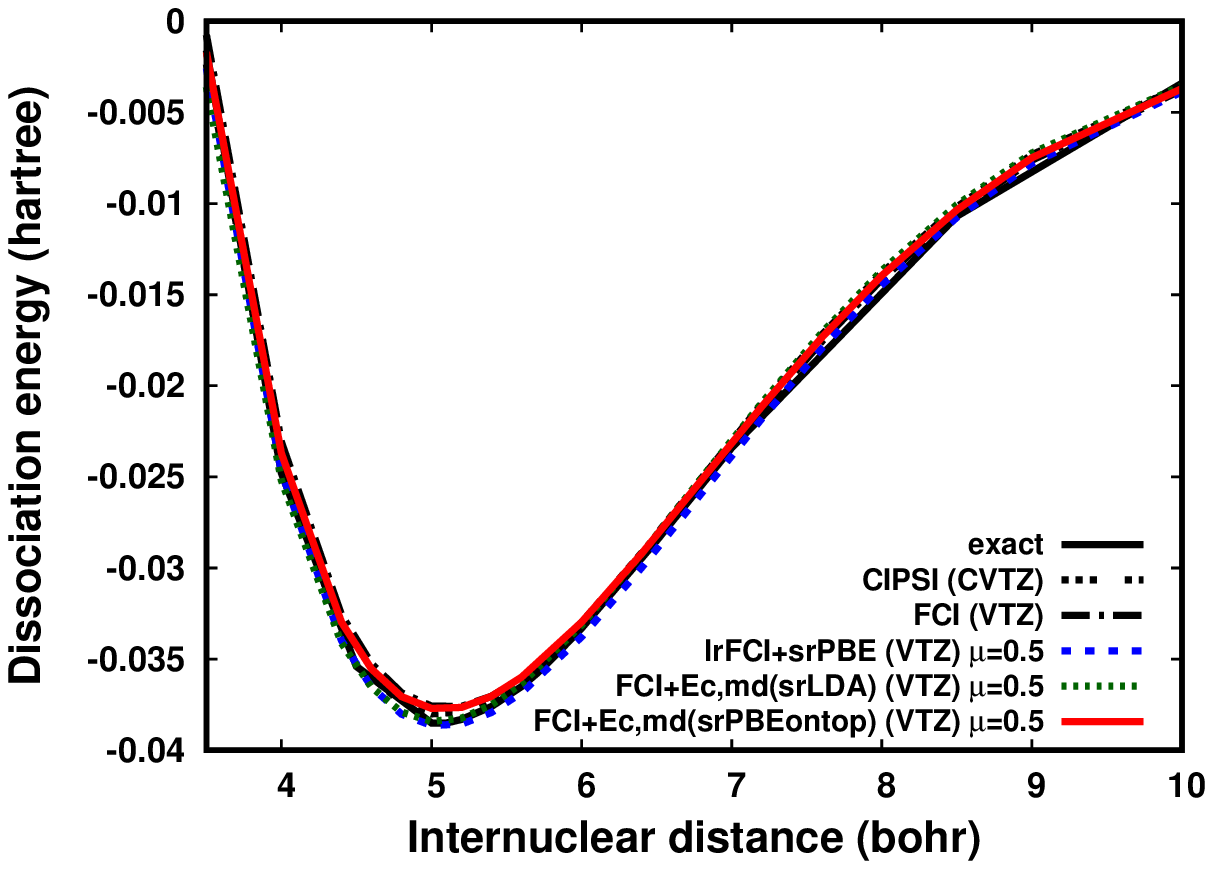}     
   \end{minipage}
   \caption{Left panel: Total energy curves of Li\tsb{2} as a function of the internuclear distance $R$ calculated by lrFCI+srPBE, FCI+E\tsb{c,md}(srLDA), and FCI+E\tsb{c,md}(srPBEontop) with a range-separation parameter of $\mu=0.5$ bohr$^{-1}$ using the cc-pVTZ (VTZ) basis set without core excitations. For comparison, the estimated exact non-relativistic energy curve~\cite{LieCle-JCP-74b} as well as the energy curves calculated by frozen-core FCI with the cc-pVTZ basis set and by a well-converged variational CIPSI with the cc-pCVTZ (CVTZ) basis set and allowing core excitations are also reported. Right panel: Dissociation energy curves, $E(R)-E(R\to\infty)$, where all the curves have been shifted so that the energy at dissociation is set to 0.}
   \label{dissociation_Li2}
\end{figure*}

\begin{figure*}[t]
   \begin{minipage}[c]{.46\linewidth}
      \centering
      \includegraphics[scale=0.6]{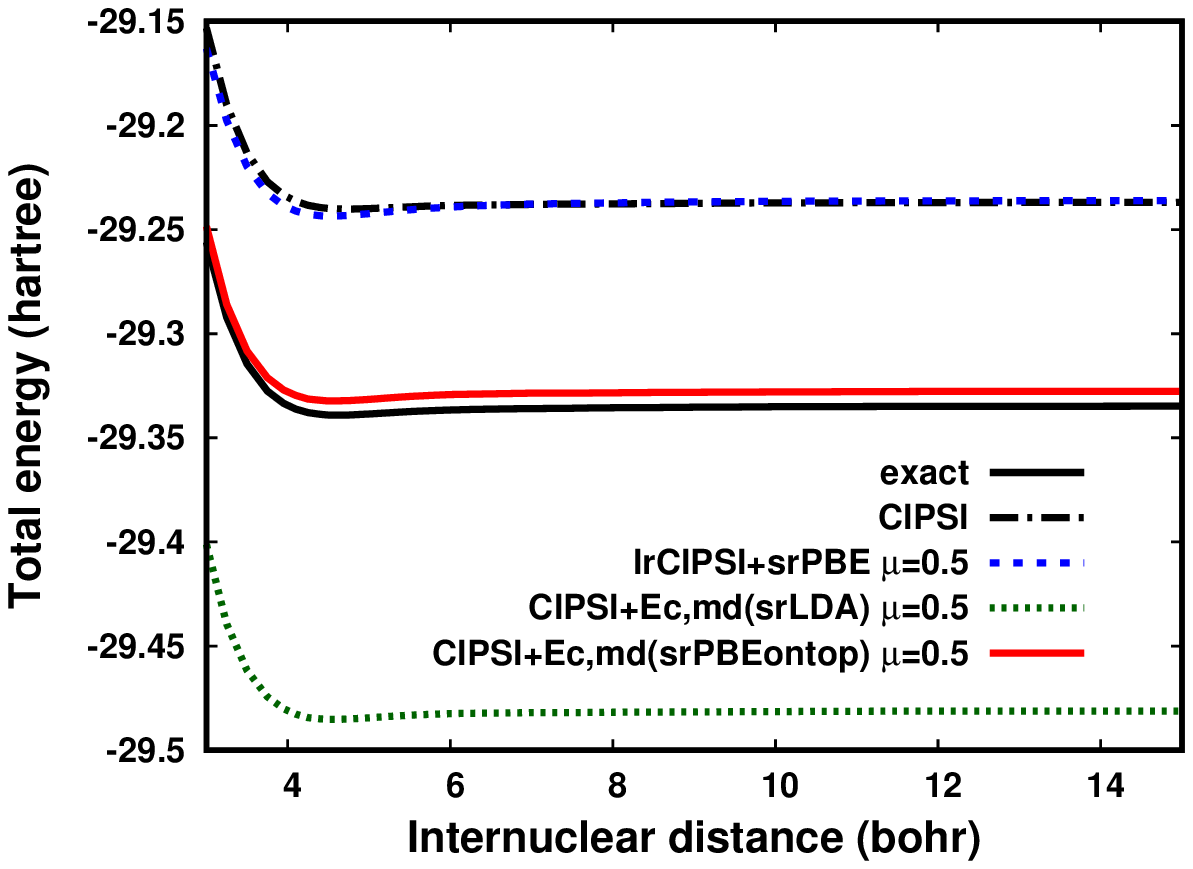} 
   \end{minipage} \hfill
   \begin{minipage}[c]{.46\linewidth}
      \centering
      \includegraphics[scale=0.6]{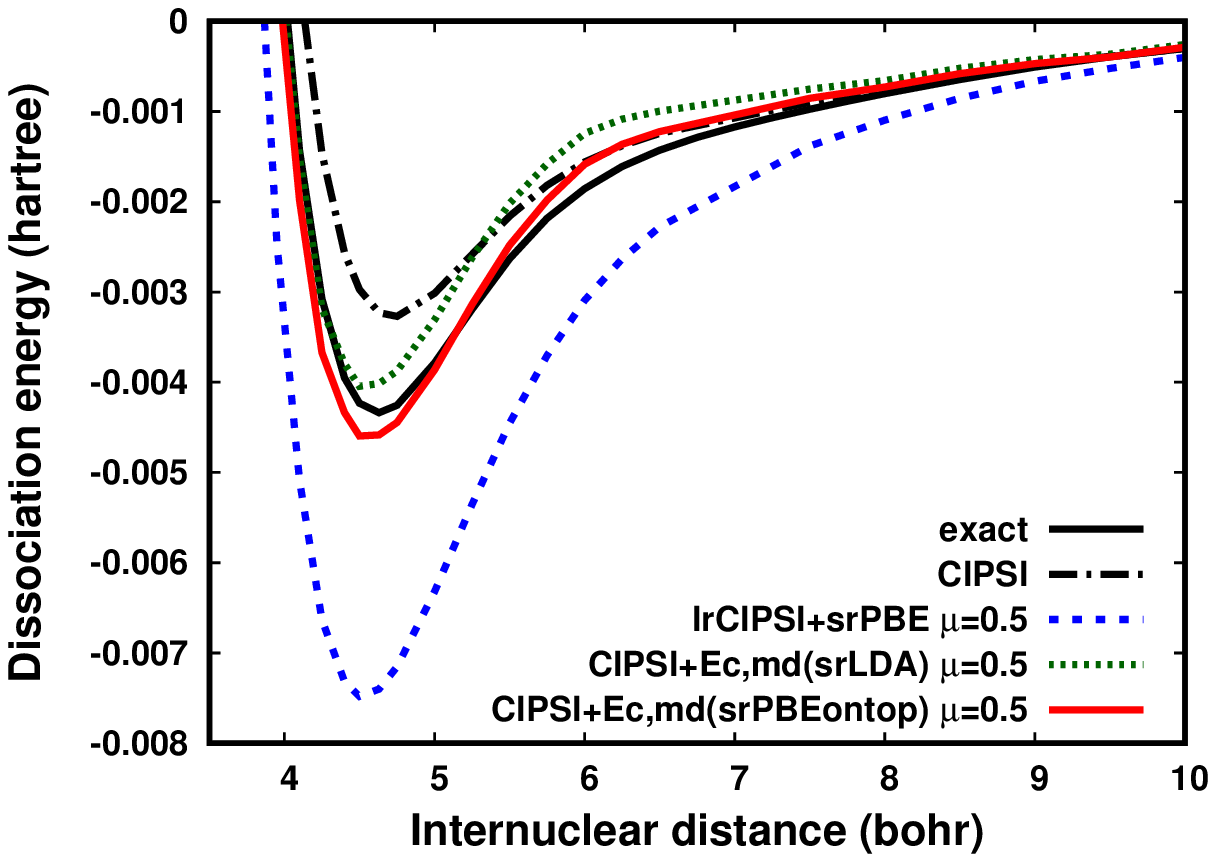}     
   \end{minipage}
   \caption{Left panel: Total energy curves of Be\tsb{2} as a function of the internuclear distance $R$ calculated by lrCIPSI+srPBE, CIPSI+E\tsb{c,md}(srLDA), and CIPSI+E\tsb{c,md}(srPBEontop) with a range-separation parameter of $\mu=0.5$ bohr$^{-1}$ using the aug-cc-pVTZ basis set without core excitations. For comparison, the estimated exact non-relativistic energy curve~\cite{RoeVes-IJQC-05} as well as the energy curve calculated by a well-converged variational frozen-core CIPSI with the aug-cc-pVTZ are also reported. Right panel: Dissociation energy curves, $E(R)-E(R\to\infty)$, where all the curves have been shifted so that the energy at dissociation is set to 0.}
   \label{dissociation_Be2}
\end{figure*}

The dissociation energy curves of the homonuclear diatomic molecules H$_2$, Li$_2$, and Be$_2$ are reported in Figs.~\ref{dissociation_H2}, \ref{dissociation_Li2}, and~\ref{dissociation_Be2}, respectively. These molecules cover different types of bonding and correlation effects. The RS-DFT calculations were performed at the frozen-core FCI level using the cc-pVTZ basis set for H\tsb2 and Li\tsb2 and at the frozen-core CIPSI level using the aug-cc-pVTZ basis set for Be\tsb{2}. We did not attempt to find an optimal value for the range-separation parameter and we simply used the common value of $\mu=0.5$ bohr$^{-1}$~\cite{GerAng-CPL-05a}. We did not try to remove the basis-set superposition error (BSSE) in the dissociation energy curve of the weakly bound Be\tsb{2} molecule since the BSSE is known to be small for this system for frozen-core calculations with triple-zeta basis sets, and even more so with range separation~\cite{GerAng-CPL-05b,ReiTouSav-TCA-18}.

\subsubsection{H\tsb2 molecule}

The electronic ground-state of the H\tsb2 molecule is one of the standard toy model of quantum chemistry owing to the range of correlation effects that it presents, from dynamic correlation at the equilibrium internuclear distance to static correlation at dissociation.

\begin{figure}[t]
        \centering
        \includegraphics[scale=0.6]{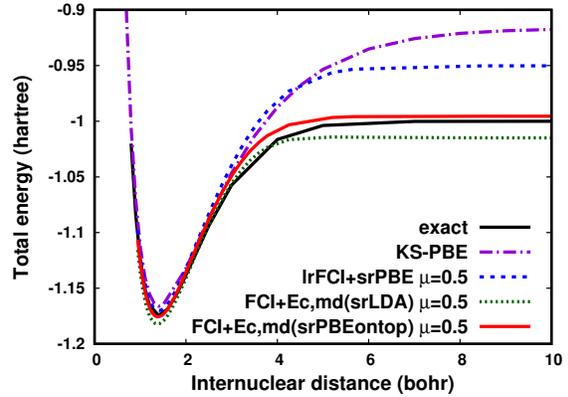}     
        \caption{Total energy curve of H\tsb{2} as a function of the internuclear distance calculated by lrFCI+srPBE, FCI+E\tsb{c,md}(srLDA), and FCI+E\tsb{c,md}(srPBEontop) with a range-separation parameter of $\mu=0.5$ bohr$^{-1}$ and the cc-pVTZ basis set. For comparison, the KS-PBE energy curve calculated with the same basis set and the estimated exact non-relativistic energy curve~\cite{LieCle-JCP-74a} are also reported.}
        \label{dissociation_H2}
\end{figure}

The KS-PBE total energy curve showed in Fig.~\ref{dissociation_H2} is a good example of the success and failure of standard KS-DFT with semilocal density approximations. KS-PBE gives an accurate energy near the equilibrium which illustrates the fact that the PBE approximation correctly describes dynamic correlation effects. By contrast, the KS-PBE results are far from being satisfying near the dissociation, which shows the incapacity of the PBE approximation to deal with static correlation effects.

The lrCI+srPBE method provides a way to partly correct the description of static correlation. Indeed, the long-range wave function accounts for part of the static correlation, while the srPBE functional accounts for the dynamic correlation. Thus, near the equilibrium, the lrCI+srPBE energy curve is essentially as accurate as the KS-PBE one (in fact slightly more accurate), and in the dissociation limit lrCI+srPBE greatly improves upon KS-PBE by giving an energy which correctly saturates. However, for the value of the range-separation parameter used ($\mu=0.5$ bohr$^{-1}$), a substantial part of the electron-electron interaction is still taken into account via the short-range Hartree-exchange-correlation functional which leads to the important remaining error at dissociation.

The CI+E\tsb{c,md}(srLDA) total energy curve is below the exact energy curve. This is of course due to the overestimation (in absolute value) of the short-range multideterminant correlation energy ${{\bar{E}_\mathrm{{c,md}}^{\mathrm{sr,}{\mu}}}}$ by the srLDA correlation functional. Still, the CI+E\tsb{c,md}(srLDA) approach constitutes for this system a substantial improvement over KS-PBE and lrCI+srPBE, especially in terms of the relative shape of the dissociation curve.

We see that CI+E\tsb{c,md}(srPBEontop) provides by far the most accurate total energy curve, either in terms of absolute energy or relative shape. The H\tsb2 molecule is simple enough to easily understand why our new srPBEontop functional gives accurate results. At dissociation, the two electrons are so far away from each other that the electron-electron interaction becomes negligible. Therefore, the exact long-range interacting Hamiltonian of RS-DFT $\hat{H}^\mu$ [Eq.~\eqref{H_mu}] becomes equivalent to the physical Hamiltonian $\hat{H}$, and consequently the exact long-range wave function $\Psi^\mu$ reduces to the exact ground-state wave function $\Psi$ of the system. Hence, at dissociation, the term $\braket {\Psi^\mu|\hat{H}|\Psi^{\mu}}$ in Eq.~\eqref{E-ecmd} should be equal to the exact energy, and the short-range multideterminant correlation energy ${{\bar{E}_\mathrm{{c,md}}^{\mathrm{sr,}{\mu}}[n]}}$ should vanish. This exact behavior is correctly recovered thanks to the on-top pair density. Indeed, at dissociation, the on-top pair density, $n_2(\textbf{r},\textbf{r})$ or $n_2^\mu(\textbf{r},\textbf{r})$, goes to zero since the two electrons are far away from each other, and it is easy to check from Eqs.~\eqref{e_srPBEontop} and~\eqref{beta_srPBEontop} that this makes the srPBEontop correlation energy vanish. We thus see that the dependence on the on-top pair density is the key to obtain the correct dissociation limit. The reason why in practice CI+E\tsb{c,md}(srPBEontop) still gives a small error at dissociation is that the short-range potential $\hat{\bar{V}}_{\mathrm{Hxc}}^{\mathrm{sr},\mu}[n]$ in Eq.~\eqref{H_mu} does not exactly vanish at dissociation due to the use of the PBE approximation. Consequently, the long-range wave function $\Psi^{\mu}$ does not exactly reduce to the exact wave function in the dissociation limit but is a good approximation to it.

\subsubsection{Li\tsb2 and Be\tsb2 molecules}

We now consider the electronic ground-state energy curves of the Li\tsb2 and Be\tsb2 molecules. Although they are still relatively small systems, they raise more difficulties than H\tsb2, not only because of the increasing number of electrons but especially because of the more subtle mix between dynamic and static correlations that has to be described. Also, these two molecules are characterized by two different types of bond: Li\tsb2 is a strongly bonded molecule, while Be\tsb2 is a weakly bonded molecule with a very shallow well of only a few millihartree. 

Here the distinction between dynamic correlation at the equilibrium geometry and static correlation at dissociation that exists in H\tsb2 is no longer valid. Static correlation effects are present at all internuclear distances for these molecules, which limits the accuracy of lrCI+srPBE total energies. Indeed, as shown in the left panels of Figs.~\ref{dissociation_Li2} and~\ref{dissociation_Be2}, the lrCI+srPBE total energy is well above the exact energy for all internuclear distances. Using the short-range multideterminant correlation approach, we obtain results following the same trends observed for H\tsb2. For both Li\tsb2 and Be\tsb2, the CI+E\tsb{c,md}(srLDA) total energy curve is far below the exact one, while the CI+E\tsb{c,md}(srPBEontop) total energy curve is quite close to the exact one.

We note that, in the case of Li\tsb2, the FCI+E\tsb{c,md}(srPBEontop) total energy curve, calculated with the cc-pVTZ basis set without core excitations, is much more accurate than the frozen-core FCI total energy curve calculated with the same basis set, and even slightly closer to the exact energy curve than the CIPSI total energy curve calculated with the cc-pCVTZ basis set and allowing core excitations. This is so because, in the FCI+E\tsb{c,md}(srPBEontop) method, core correlation being a short-range effect is included in the \mbox{srPBEontop} functional. Thus, RS-DFT allows one to drop core excitations in the expansion of the wave function without losing accuracy, which is another important advantage in terms of computational cost.

We now discuss the relative dissociation energy curves, $E(R)-E(R\to\infty)$ where $R$ is the internuclear distance, shown in the right panels of Figs.~\ref{dissociation_Li2} and~\ref{dissociation_Be2}. For Li\tsb2, all the methods tested here give almost the same relative dissociation energy curve and is very close to the exact relative energy curve. Thus, even though these methods give very different total energies, they all provide an accurate estimation of both the equilibrium distance and the dissociation energy.

For Be\tsb2, the different methods give more diverse relative dissociation energy curves. This is due to the fact that we are looking at a much smaller energy scale in comparison to Li\tsb2, and also to the fact that the Be\tsb2 bond involves a complex mix of correlation effects. The full-range frozen-core CIPSI calculation using the aug-cc-pVTZ basis set gives a substantially underestimated dissociation energy and a slightly overestimated equilibrium distance, i.e. it favors too much the separated atoms over the more correlated bonded molecule. This is due to the incompleteness of the basis set and possibly also to the missing of core excitations. On the contrary, lrCIPSI+srPBE largely overestimates the dissociation energy and slightly underestimates the equilibrium distance, i.e. it favors too much the bonded molecule over the dissociated atoms. In this case, the main source of error comes from the srPBE exchange-correlation functional (fractional-charge and/or fractional spin errors). The short-range multideterminant correlation approach gives quite good relative dissociation energy curves. The CI+E\tsb{c,md}(srLDA) relative dissociation energy curve is a bit too high, in particular at long distances. The CI+E\tsb{c,md}(srPBEontop) relative dissociation energy curve is the closest to the exact one, showing the srPBEontop functional properly accounts for differential correlation effects.

\section{Conclusion}
\label{conclusion}

In this work, we have developed a new approximation to the short-range multideterminant correlation functional ${{\bar{E}_\mathrm{{c,md}}^{\mathrm{sr,}{\mu}}[n]}}$ involved in the variant of RS-DFT given by Eq.~\eqref{E-ecmd}. This approximation, named srPBEontop, is a local functional of the density, the density gradient, and the on-top pair density, which locally interpolates between the standard PBE correlation functional at vanishing range-separation parameter $\mu$ and the known exact asymptotic expansion of the functional at large $\mu$. By combining this srPBEontop correlation functional with (selected) CI calculations for the long-range wave function, one expects to obtain a multideterminant RS-DFT method which is essentially free from self-interaction errors and appropriately accounts for both short-range dynamic correlation and static correlation. This is supported by the accurate dissociation energy curves of the small but diversely correlated molecules H$_2$, Li$_2$, and Be$_2$ that we have obtained with the multideterminant RS-DFT approach with the srPBEontop approximation.

Besides assessing the present method on more systems, possible future developments include adding the second-order CIPSI perturbative correction, performing self-consistent calculations with the srPBEontop approximation, combining this approximation with the recent local-$\mu$ approach of Ref.~\onlinecite{GinPraFerAssSavTou-JCP-18}, and calculating excited states for example using perturbation theory along the ground-state range-separated adiabatic connection~\cite{RebTouTeaHelSav-MP-15,RebTeaHelSavTou-MP-18} or using ghost-interaction-corrected ensemble RS-DFT~\cite{SenHedAlaKneFro-MP-16,AlaKneFro-PRA-16,AlaDeuKneFro-JCP-17}.

\begin{acknowledgements}
We thank Pierre-Fran{\c c}ois Loos and Cyrus Umrigar for comments on the manuscript.
\end{acknowledgements}


\end{document}